\documentclass[twocolumn,showpacs,amsmath,superscriptaddress,amssymb,prb,floatfix]{revtex4-2} 
\usepackage{amssymb,amsmath}
\usepackage{color}
\usepackage{graphicx}
\usepackage{dcolumn}
\usepackage{bm}
\usepackage{upgreek}
\usepackage{marvosym}
\usepackage{latexsym,epsfig} 
\usepackage{bm}
\pdfoutput=1 
\usepackage[T1]{fontenc}
\usepackage{hyperref}
\usepackage[nodisplayskipstretch]{setspace}

\DeclareUnicodeCharacter{2212}{-}

\begin{document}
\title{Enhancing Stability, Magnetic Anisotropy, and Coercivity of  $\tau$-L$1_0$ MnAl: Machine Learning, \textit{Ab Initio}, and Micromagnetic Modeling}
\author{Churna Bhandari}
\email{Corresponding Author: cbb@ameslab.gov}
\affiliation{The Ames National Laboratory, U.S. Department
of Energy, Iowa State University, Ames, Iowa 50011, USA}
\author{Gavin N. Nop}
\affiliation{The Ames National Laboratory, U.S. Department
of Energy, Iowa State University, Ames, Iowa 50011, USA}
\affiliation{Department of Mathematics,  Iowa State University, Ames, Iowa 50011, USA}
\author{T. Lograsso}
\affiliation{The Ames National Laboratory, U.S. Department
of Energy, Iowa State University, Ames, Iowa 50011, USA}
\author{Durga Paudyal}
\email{Current Address: {Department of Physics and Astronomy, University of Iowa, Iowa City, Iowa 52242, USA}}
\affiliation{The Ames National Laboratory, U.S. Department of Energy, Iowa State University, Ames, Iowa 50011, USA}
\affiliation{Department of Electrical and Computer Engineering,  Iowa State University, Ames, Iowa 50011, USA}

\begin{abstract} 
The binary manganese aluminium (MnAl) alloy with L$1_0$ crystal structure is a promising rare earth element-free permanent magnetic material because of its exceptional magnetic properties. However, experimentally synthesizing it in a stable bulk form is extremely challenging. Here, an alternative method of stabilizing the material is proposed and theoretically verified by partially substituting Mn and Al sites with Fe and Ni and identifying its enhanced phase stability, magnetic anisotropy, and coercivity from density functional theory (DFT), machine learning (ML) crystal graph convolution neural network (CGCNN), and micro-magnetic modeling. When considering a fixed ($50\%$)-Ni, the magnetic anisotropy increases with the increasing Fe content but decreases the formation energy. The calculated formation energies, elastic constants, and phonon frequencies demonstrate that all the binary and quaternary compositions are stable. Most importantly, the magnetic moment and magnetic anisotropy constants in $50\%$-Fe substituted composition (equiatomic phase) increase significantly compared to the MnAl. The predicted coercivity of the equiatomic phase is larger than the parent compound calculated by combining DFT computed parameters with micro-magnetic simulations.
\end{abstract}
\maketitle

\section{Introduction}
Since the early 1900s experiments on binary alloys conducted by Hindiricks\cite{Hindricks2908} and by Ishiwara 
\cite{Ishiwara1930SRTI}, there has been an ongoing pursuit to harness the critical element-free ferromagnetic equiatomic MnAl with $\tau-$L$1_0$ phase for permanent magnet applications\cite{KochJAP60, KonoJPS58, ZhangScripta94, KlemmerScripta95}. The  $\tau-$ phase with L$1_0$ type tetragonal crystal symmetry transforms from the non-magnetic hexagonal $\epsilon$-phase\cite{HoydickJAP97} under specific cooling conditions within a narrow range of ($\sim 51 - 58 \%$) Mn-stoichiometric conditions\cite{KonoJPS58,KochJAP60}.
Although it is metastable, it possesses high magnetic moment \cite{SakumaJPS94}, magnetic anisotropy\cite{ParetiJAP86,Zhao2015MagneticAO}, Curie temperature\cite{OhtaniIEE77,KochJAP60} ($T_{\text C}$= 380$^\circ$C or 653.15 K), and maximum energy product,  BH${_{\text max}}=$ 12 KJ/m$^3$ (14 MGOe)\cite{CoeyScripta012}. These modest magnetic properties enable it to be a promising gap magnet that fulfills the performance gap between low-performing ferrites \cite{BhandariPRapplied23, BhandariPhysRevResearch, BhandariPRM21} and high-performing rare-earth permanent magnets\cite{CoeyScripta012, McCallumAnnualRef014, CoeyJAP014, LuJAC016}. However, synthesizing it in bulk form is challenging, except the specific forms such as ribbons\cite{JanotovaJAC2017} and thin films\cite{HuangJAP015, Zhao2015MagneticAO, OoganeJJAS017}.
Introducing dopants\cite{KonoJPS58,ParetiJAP86,FengJAC021}, such as carbon, can improve its structural, mechanical\cite{ParetiJAP86}, and thermal\cite{XiangJAC019} stabilities, resulting in the formation of a ternary MnAlC. Nonetheless, this approach leads to a decrease in $T_{\text C}$\cite{RossiterMetal84, KonoJPS58, OhtaniIEE77, DreizlerIEE80} and anisotropy field\cite{ParetiJAP86}, including an increase in manufacturing cost due to the hot extrusion process\cite{BohlmannJAP81}.

The L$1_0$ crystal structure is commonly found in a family of ferromagnetic materials, FePd, FePt, and CoPt binaries. This structure forms due to the first-order transition from cubic (A$_1$) structure at low temperatures\cite{ZhangScripta94, KlemmerScripta95} by a nucleation and growth process. The conventional L$1_0$ crystal structure consists of ABC$_2$ arrangements\cite{SKOMSKI2005389} (4 basis atoms per unit cell, space group 123 $P4/mmm$, the Pearson symbol tP4 \cite{LaughlinScripta05}), where three sublattices A, B, and C occupy $1a$(0, 0, 0), $1c$(1/2, 1/2, 0), and $2e$(1/2, 0, 1/2) and (0, 1/2, 1/2) Wyckoff sites, respectively. As the CuAu(I) L$1_0$-type structure has only two types of basis atoms, its crystal structure can be described either by a four-atom base-centered tetragonal cell (tP4) or a two-atom tetragonal unit cell (Pearson symbol tP2 \cite{LaughlinScripta05}).
The tetragonal unit cell with lattice constants ($c^{'}=c$ and $a^{'}=a/\sqrt{2}$, where $c$ and $a$ are the lattice vectors of tP4 unit cell) is composed of two distinct basis atoms occupying $1a$(0, 0, 0) and $1d$(1/2, 1/2, 1/2) Wyckoff positions. 

In general, MnAl exhibits a stable antiferromagnetic phase when it has a high Mn content $\rm{Mn} >50\%$ \cite{AlexanderPRB014}. However, achieving the equiatomic MnAl $\tau-$phase is not straightforward as excess Mn atoms tend to occupy Al sites, easily decomposing into equilibrium phases, including $\gamma_2$-Mn$_5$Al$_8$ (space group $R3m$) and $\beta$-Mn (space group $P4_{132}$)\cite{ManchandaJAP015}.
When the Mn content decreases gradually, it transitions into a ferromagnetic state exhibiting a strong Mn-dependent Curie temperature\cite{ZengJMM07}. MnAl forms a ferromagnetic metastable $\tau-$ phase, a nominal L$1_0$ CuAu(I)-type tetragonal structure, through the cooling of the high-temperature $\epsilon$-phase, with a space group of $P6_3/mmc$\cite{KonoJPS58} or through annealing\cite{MakinoJap63}. In a separate experiment, Klemmer {\sl et al.} reported that the $\tau$ phase emerges via shear or displacive transformation\cite{KlemmerScripta95}.
However, it is prone to issues such as large grain size, antiphase boundaries, and twin-like defects\cite{LanduytIEE78, BittnerActa015}, which negatively affect its magnetic properties.

To improve the stability and magnetic properties of MnAl alloys, various elements such as Ti, Cu, Ni, C, B\cite{Sakka1989}, Fe\cite{ManchandaJAP015}, Co\cite{PaduaniIntermetallics010}, Ni\cite{MicanJMM016}, Cu\cite{SaravananAPL015}, Zn\cite{WangJM011}, and rare earth elements (Pr and Dy)\cite{Liu2012} have been substituted into the MnAl alloy. In a study by Saravanan {\sl et al.}\cite{SaravananAPL015}, Cu-substituted MnAl exhibited exceptionally high coercivity (H$_\text{c}$). C and Zn have been shown to enhance the stability, coercivity, and saturation magnetization of the $\tau$-phase\cite{OhtaniIEE77, WangJM011}, while Fe alone has been predicted to enhance the magnetic anisotropy\cite{ManchandaJAP015}. However, although C stabilizes the $\tau$ phase, it does not prevent the formation of equilibrium phases during annealing\cite{ZengJMM07}. In this work, an alternative approach is proposed to enhance stability and magnetic properties by introducing Fe and Ni simultaneously.

In this paper, a first-principles study is presented for the stability, electronic properties, magnetic properties, and $H_c$ along with micromagnetic modeling of pristine and site substituted MnAl alloys. The phase stabilities are investigated by calculating formation energy, mechanical properties, and phonon dispersions. The site substituted MnAl alloys are stable compared to other unary and potential binary phases.  The formation energy and saturation magnetization density are explored by machine learning (ML) study on density functional theory (DFT) calculated magnetic materials. The calculations of elastic properties confirm the mechanical stability of all compositions. The absence of soft-mode frequencies in the phonon dispersion indicates their dynamic stability. The introduction of dopants leads to changes in the electronic structure, affecting the magnetic moments and anisotropy. In particular, improvements are predicted in the magnetic anisotropy and coercivity for the equiatomic phase consisting of equal amounts of Fe, Mn, Al, and Ni obtained by simultaneous doping with Fe and Ni.

\section{Crystal structure and Methodology}
\begin{figure}
\includegraphics[scale=0.21]{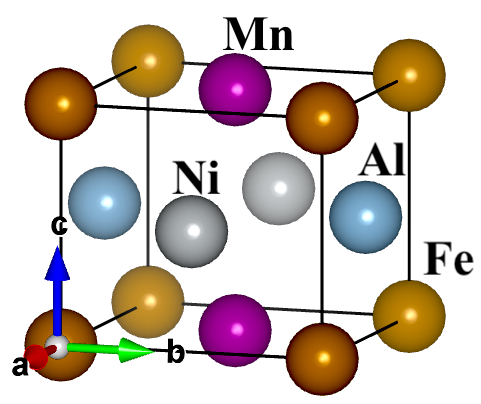}
\includegraphics[scale=0.225]{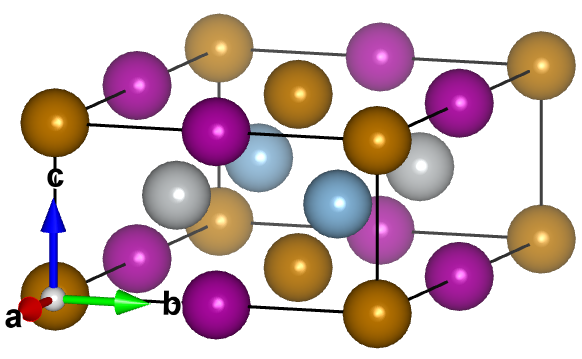}
\caption{The simulated crustal structures of site substituted MnAl i.e., FeMnAlNi ABC$_2$ L1$_0$ type distorted tetragonal ($Pmmm$ space group no. 47) {\it left} and monoclinic  crystal with space group no. 10 derived from the tetragonal $\tau-$L$1_0$ crystal of metastable MnAl with space group no. 123 $P4/mmm$ {\it right}.}
    \label{fig:structure}
\end{figure}
The Vienna simulation package (VASP) was used to solve Kohn-Sham eigenvalue equations with generalized gradient approximation Perdew-Burke-Ernzerhof (PBE) functional\cite{PBE, PBE1} within the projector augmented wave method (PAW)\cite{KressePRB93, KressePRB99}. The structure was relaxed using a plane-wave energy cutoff of 520 eV and a $9\times9\times7$ {\bf k}-mesh for Brillouin zone integration in MnAl with similar {\bf k}-meshes for site substituted-compositions proportional to their supercell lattice constants, including for distorted tetragonal L1$_0$ structure. To obtain the magnetocrystalline anisotropy (MCA), fully self-consistent relativistic calculations were performed, including the spin-orbit interaction as implemented in the VASP \cite{KressePRB00}. To achieve the convergence of the magnetic anisotropy energy, several calculations were performed varying {\bf k}-mesh up to $14\times 14\times 13$ for MnAl, $10\times10\times 10$ for FeMnAlNi, and $7\times10\times 19$ for other partially site substituted systems commensurate with lattice constants. 

For the site substituted-compositions, the two lowest total energy structures were used among the considered crystal structures (i) a 4 atom standard L1$_0$ tP4 unit cell for FeMnAlNi as shown in Fig. \ref{fig:structure} {\it left} and (ii) an 8 atom new supercell Fig. \ref{fig:structure} {\it right} . In the L1$_0$ unit cell, one Mn was replaced by Fe and one Al by one Ni, and the resulting structure has the same proportion of four elements, which is considered as an equiatomic phase. This crystal structure with space group no. 47 $Pmmm$ is spanned by Fe, Mn, Al, and Ni atoms occupying $1a$, $1f$, $1d$, and $1g$ Wyckoff sites. 
For structure (ii), a $2\times2\times1$ supercell was built from $\tau-$ L$1_0$-type (tP2 unit cell) structure by partially replacing Mn and Al with Fe and Ni. For FeMnAlNi, a primitive cell with space group no. 10 from the supercell was built, which is utilized in the calculations. The crystal structures are spanned by the four inequivalent sublattices Fe, Mn, Al, and Ni at $1a$, $1g$, $1e$, and $1f$ Wyckoff positions in FeMnAlNi. However, as discussed later in Section \ref{formationenergy}, it is noted that the total energy is higher for this structure. For FeMnAlNi, only the L$1_0$ tP4 type unit cell (i) results are shown. The unit cell (ii) is used for the partially ($25\%$) Fe-and $50\%$ Ni-substituted composition (Fe$_{0.5}$Mn$_{1.5}$AlNi). The crystal structure of Fe$_{0.5}$Mn$_{1.5}$AlNi is spanned by the four-six inequivalent sublattices Fe at $1a$, Mn$_1$ at $1g$, Mn$_2$ at $1c$, Mn$_3$ at $1d$, Al at $2n$, and Ni at $2n$ Wyckoff positions. The same supercell is used for the Ni-only substituted case, Mn$_2$AlNi, but it has a different space group, no. 65 occupied by three sublattices, Mn, Al, and Ni, at $4a$, $2c$, and $2d$ Wyckoff positions.

\begin{table}
    \begin{ruledtabular}
      \caption{Optimized lattice parameters of pristine and Fe/Ni-substituted MnAl in units of \AA~ and comparison with available experiments.}    \label{tab1}
    \begin{tabular}{l l l l}
  System &$a$ & $b$ & $c$\\
  \hline
  MnAl &2.76 &  2.76 & 3.46  \\
  Expt. &2.77\cite{KinemuchiRSC016, KochJAP60} & 2.77 &3.57\cite{KochJAP60},  3.59\cite{KinemuchiRSC016}\\
   & 2.75\cite{OskarJMR71} &2.75  & 3.56\cite{OskarJMR71}\\
   &2.769\cite{DreizlerIEE80} &2.769  & 3.618\cite{DreizlerIEE80} \\
 FeMnAlNi\footnote{Tetragonal 4 atom unit cell} & 3.37 & 3.79 & 3.81\\
Fe$_{0.5}$Mn$_{1.5}$AlNi & 7.18 &5.18 &2.69\\
Mn$_2$AlNi & 7.12 & 5.24 & 2.73\\
    \end{tabular}
    \end{ruledtabular}
\end{table}

Optimized lattice constants are presented in Table \ref{tab1}. For MnAl, the PBE functional produces in-plane lattice constants very close to the experiment, although it underestimates the out-of-plane lattice constant ($c$). The out-of-plane lattice constant reduces by $\sim 3\%$ contrary to the GGA functional, which generally yields $1-2\%$ larger for other materials. However, it is noted that the experimental lattice constants are obtained in different measurement conditions. For instance, Kinemuchi {\sl et al.} measured at high pressure ($> 5$ GPa) \cite{KinemuchiRSC016}. Koch {\sl et al.}, on the other hand, found slightly different values under cooling at an average rate of 30$^{\circ}$/sec at higher temperature\cite{KochJAP60}. The lattice constants decrease with increasing Fe, as expected, because its atomic size is smaller than Mn.  



The FM state with all atoms spins except Al is parallel and has a lower energy by about 0.33 eV than the AFM structure for FeMnAlNi.
Other compositions are also FM.  
In Mn$_2$AlNi, FM-aligned Mn-spins have lower energy by 19 meV/atom relative to the AFM compared to Fe-doped compositions. These results suggest that as soon as Ni is introduced, the  L1$_0$-$\tau$ state continues to be the FM.
It is noted that Mn$_2$AlNi has heusler\cite{Heusler1903} type face-centered cubic crystal structure (space group no. 216), which is an AFM.\\


\section{Phase stability}
\subsection{Formation energy from DFT}\label{formationenergy}
The formation energy is examined relative to i) unaries and ii) binaries MnAl, FeNi, AlNi, and MnFe. 
The formation energy of Fe$_2$Mn$_2$Al$_2$Ni$_2$ (FeMnAlNi) with respect to the unaries is determined using the equation:
\begin{equation}
E_f[\text{Fe}_2\text{Mn}_2\text{Al}_2\text{Ni}_2] = E[\text{Fe}_2\text{Mn}_2\text{Al}_2\text{Ni}_2] - \sum_i n_i E_i,\label{eq1}
\end{equation}
where $E_f$ represents the formation energy, $E$[Fe$_2$Mn$_2$Al$_2$Ni$_2$] and $E_i$ denote the energies of Fe$_2$Mn$_2$Al$_2$Ni$_2$ and the $i^{th}$ element, and $n_i$ represents the number of $i^{th}$-element present in the unit cell. With respect to the binaries, the formation energy is calculated as:
\begin{eqnarray}
E_f[\text{Fe}_2\text{Mn}_2\text{Al}_2\text{Ni}_2]& =& E[\text{Fe}_2\text{Mn}_2\text{Al}_2\text{Ni}_2] \nonumber \\
& &- 2E[\text{MnAl}) -2E(\text{FeNi}],~~\label{eq2}
\end{eqnarray}
where $E$[MnAl] and $E$[FeNi] correspond to the energies of MnAl and FeNi, respectively.

\begin{figure}
\includegraphics[scale=0.35]{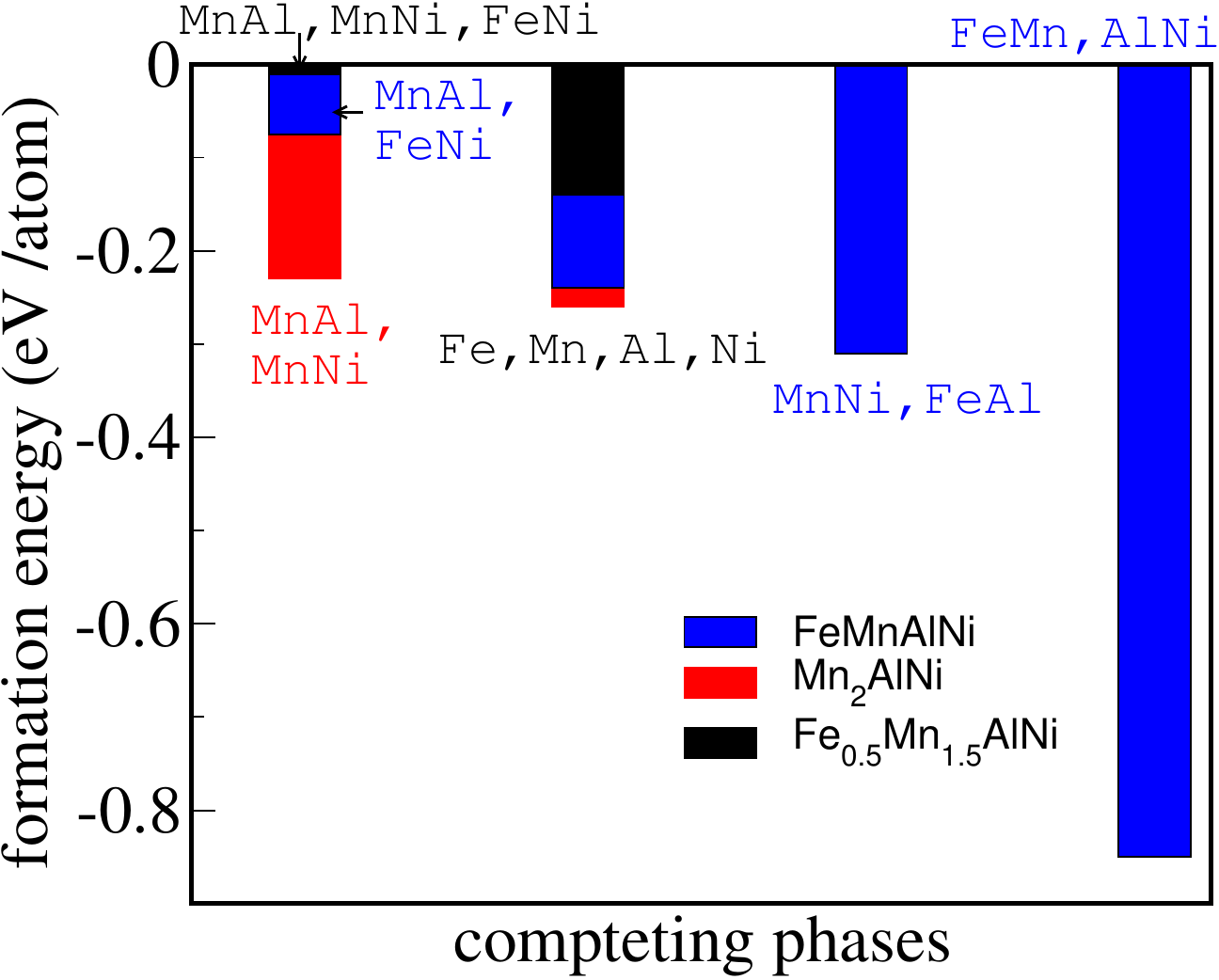}
\caption{Formation energy of Fe/Ni-doped MnAl alloy as a function of various binaries and unaries. The formation energy is negative in all phases. The 50\%Fe and 50\%Ni-doped composition has distorted L$1_0$ type structure.}
    \label{fig:formation}
\end{figure}

In both cases, the computed formation energies for all compositions are negative, as shown in Figure \ref{fig:formation}.
As discussed earlier, 50\%-Fe- and Ni-doped alloys can form two probable structures. However, the total energy of tetragonal FeMnAlNi is lower by 0.33 eV/f.u. than in a monoclinic structure. In fact, the monoclinic structure has positive formation energy (9 meV/atom) with respect to MnAl and FeNi binaries, suggesting it is an energetically unstable structure.
The Mn$_2$AlNi exhibits a similar negative formation energy. This is the heusler-type face-centered cubic crystal structure \cite{Heusler1903, Saal2013}, which is antiferromagnetic (AFM) and therefore is not important for permanent magnet applications.


\subsection{Mechanical stability}
The energy strain relation was employed to compute the mechanical properties, which, in Voigt notation, reads 
\begin{equation}
    E= \dfrac{V_0}{2}\sum_{i=1}^6\sum_{j=1}^6C_{ij}\epsilon_i\epsilon_j,
\end{equation}
where $V_0$ is the volume of unit cell, $C_{ij}$ are the elastic constants, and $\epsilon_i$ are components of strain tensor.
The calculated elastic stiffness matrix [Appendix Eq. (A1)] for MnAl is definite positive, has all positive eigenvalues, and its elements
satisfy the mechanical stability criteria for tetragonal structure\cite{MouhatPRB014} $$C_{11} > |C_{12}|, ~2C_{13}^2 < C_{33}(C_{11} + C_{12}), {\text{~and~}}
 C_{44} > 0.$$

\begin{table}
\begin{ruledtabular}
\caption{Bulk Modulus (K), Shear Modulus (G), Young's Modulus (E), Poisson's Ratio ($\nu$) in GPa and Poisson and bulk to shear modulus ratio of MnAl in Voigt notations.}\label{modulii}
\begin{tabular}{l l l l }
 parameters &   MnAl   &   FeMnAlNi   &    Fe$_{0.5}$Mn$_{1.5}$AlNi   \\
\hline
K     &  132.62  &  139.60  &  128.75   \\
 G    &   98.34  &   80.08  &   80.68  \\
 E  &  236.55  &  201.68  &  200.22  \\
 $\nu$    &    0.20  &    0.26  &    0.24  \\
   K/G         &    1.35  &    1.74  &    1.60  
 \end{tabular}
 \end{ruledtabular}
 \end{table}

Here, the elastic stiffness matrices are definite positive and satisfy the criteria for mechanical stability.
For the distorted tetragonal structure, mechanical criteria\cite{WuPRB07} $C_{11} > 0$,
 $C_{11}*C_{22} > C_{12}^2$,
$C_{11}*C_{22}*C_{33} + 2C_{12}*C_{13}*C_{23} - C_{11}*C_{23}^2 - C_{22}*C_{13}^2 - C_{33}*C_{12}^2 > 0$,
 $C_{44} > 0$,
$C_{55} > 0 $, and
$C_{66} > 0$ are satisfied.

The bulk modulus is larger in the substituted alloys, suggesting a slight increase in stiffness, while the shear modulus is the opposite with respect to pristine composition. Overall, both compositions show similar elastic properties. Similarly, Fe$_{0.5}$Mn$_{1.5}$AlNi and Mn$_2$AlNi are found mechanically stable as the elastic constants presented in \ref{stiffFexMn3x} and \ref{stiffMn2} satisfy the stability criteria given in Ref.~(\cite{WuPRB07}).
Fe lowers the stiffness at 25\% and increases slightly at 50\% (\ref{modulii}).

\subsection{Phonon dynamics}
The frozen-phonon method\cite{frozen-phonon} was used to compute the phonon frequencies. The calculations were performed with the phonopy package\cite{phonopy}, building $4\times4\times 4$ supercells for pristine materials and $2\times2\times 4$ supercells for doped compounds. For FeMnAlN, distorted tetragonal structure, a $3\times3\times3$ supercell was used. The dynamical matrix was computed using a well-converged {\bf k-} point mesh, which was further diagonalized to produce phonon frequencies with phonopy.
\begin{figure}
\includegraphics[scale=0.375]{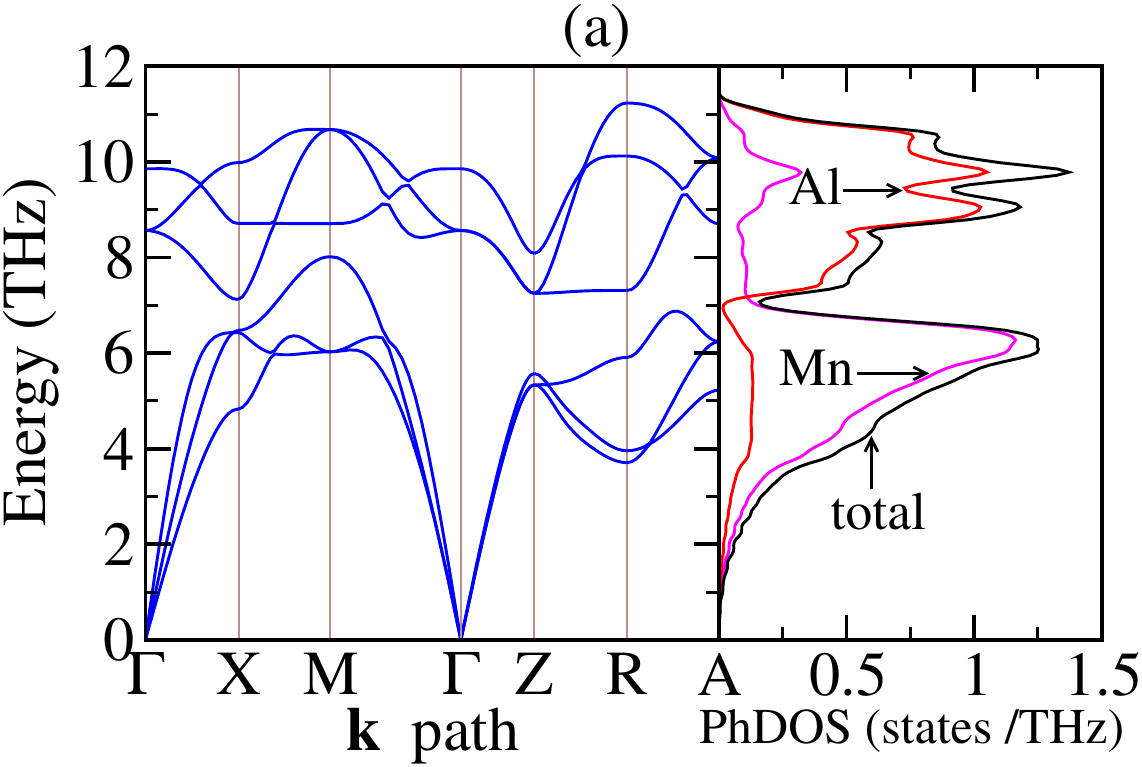}
\includegraphics[scale=0.375]{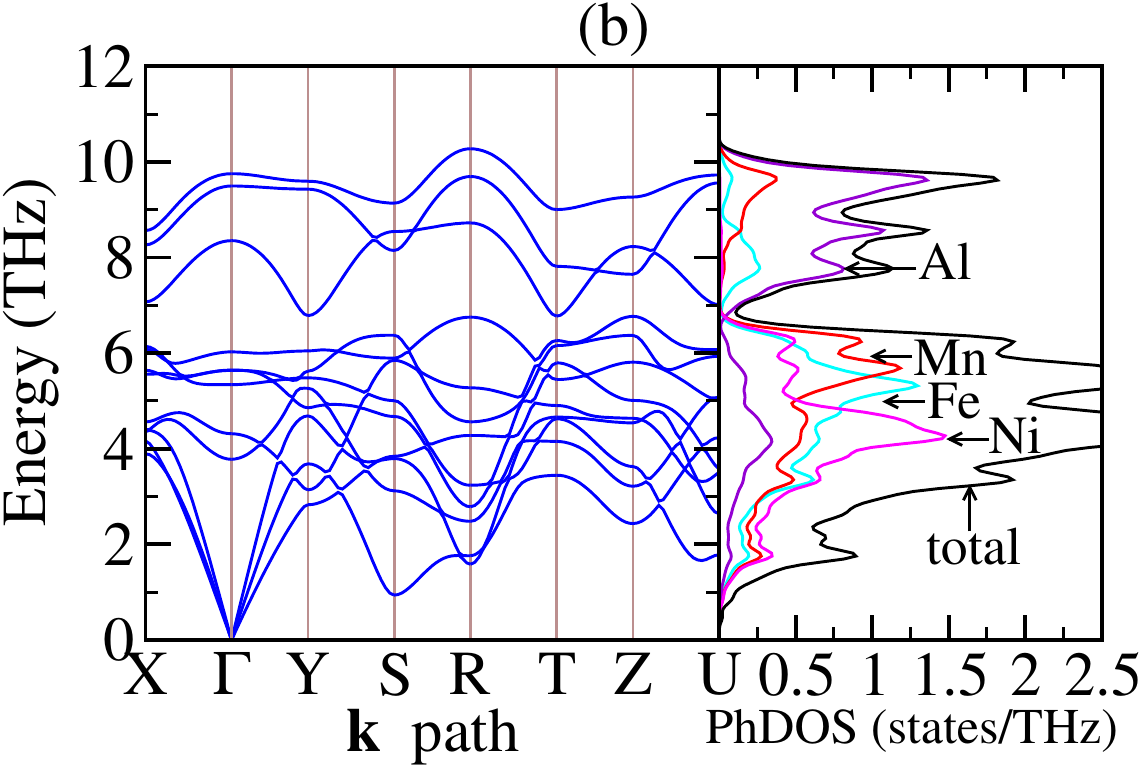}
\includegraphics[scale=0.375]{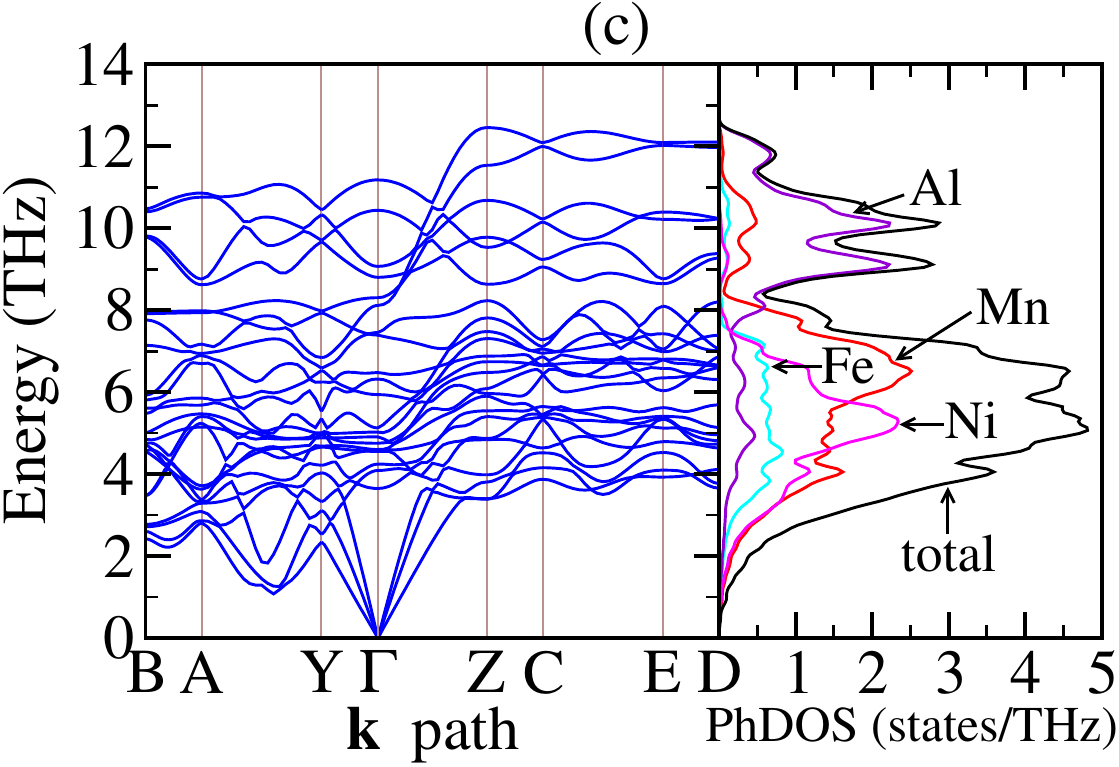}
\includegraphics[scale=0.375]{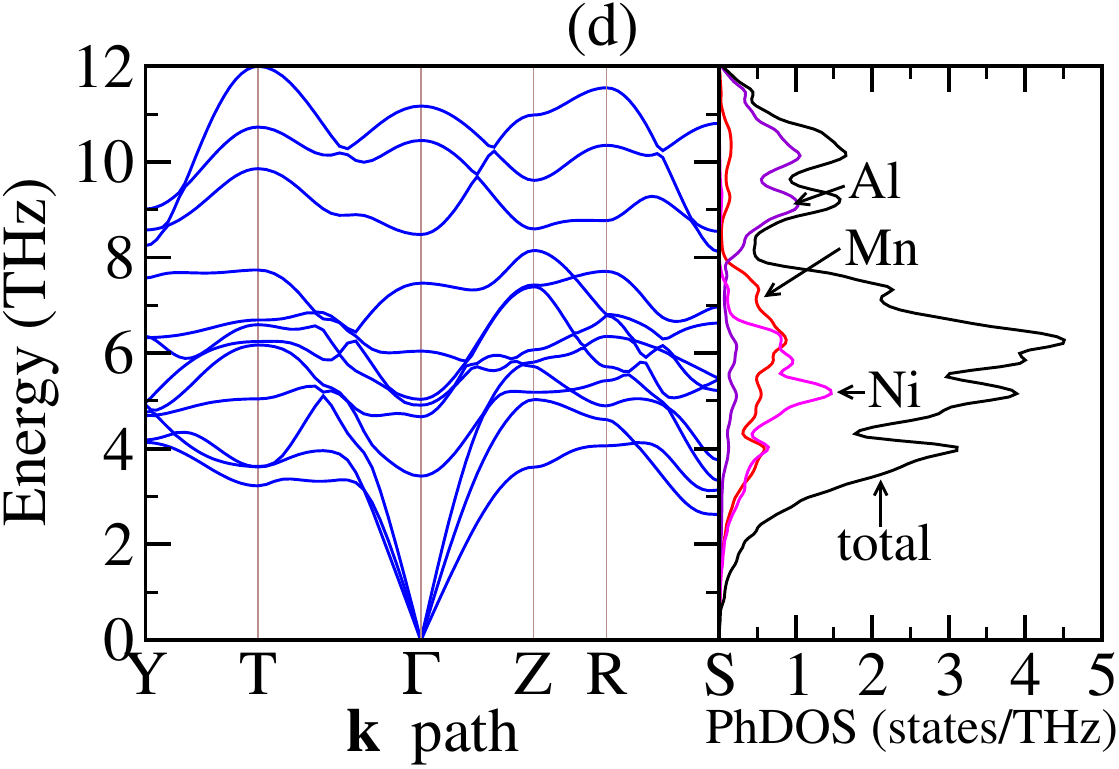}
\caption{Phonon band structure and density states for (a) MnAl, (b) FeMnAlNi, (c) Fe$_{0.5}$Mn$_{1.5}$AlNi, and (d) Mn$_2$AlNi. Total PhDOS is given per formula unit (f.u.) cell. In both cases, phonon frequencies are positive indicating their dynamical stability.}
    \label{fig:phonon}
\end{figure}
Figure \ref{fig:phonon} shows the phonon dispersion along the high symmetry {\bf k-} points for (a) pristine and (b) - (d) doped-MnAl, obtained through GGA calculations. 

Figure \ref{fig:phonon} (b) shows the phonon dispersion for distorted tetragonal FeMnAlNi.
The absence of imaginary phonon frequencies in doped and pristine compositions ensure their dynamical stability. Analyzing the atom-projected phonon density of states for MnAl, it becomes evident that Mn-derived states dominate the low-energy sector, while Al-derived states dominate the high-energy sector, consistent with their respective atomic masses. A similar trend is observed in the doped compositions, including Fe/Ni-derived states that exhibit substantial hybridization in the low-energy sector due to their closely matched atomic masses. Notably, the density of states shift upwards as the atomic masses decrease, following the sequence Mn$<$Fe$<$Ni.
Moreover, introducing Fe in the doped compositions leads to the lifting of phonon mode degeneracy due to the reduction in crystalline symmetry.
\section{Electronic structure}
\begin{figure}
\includegraphics[scale=0.275]{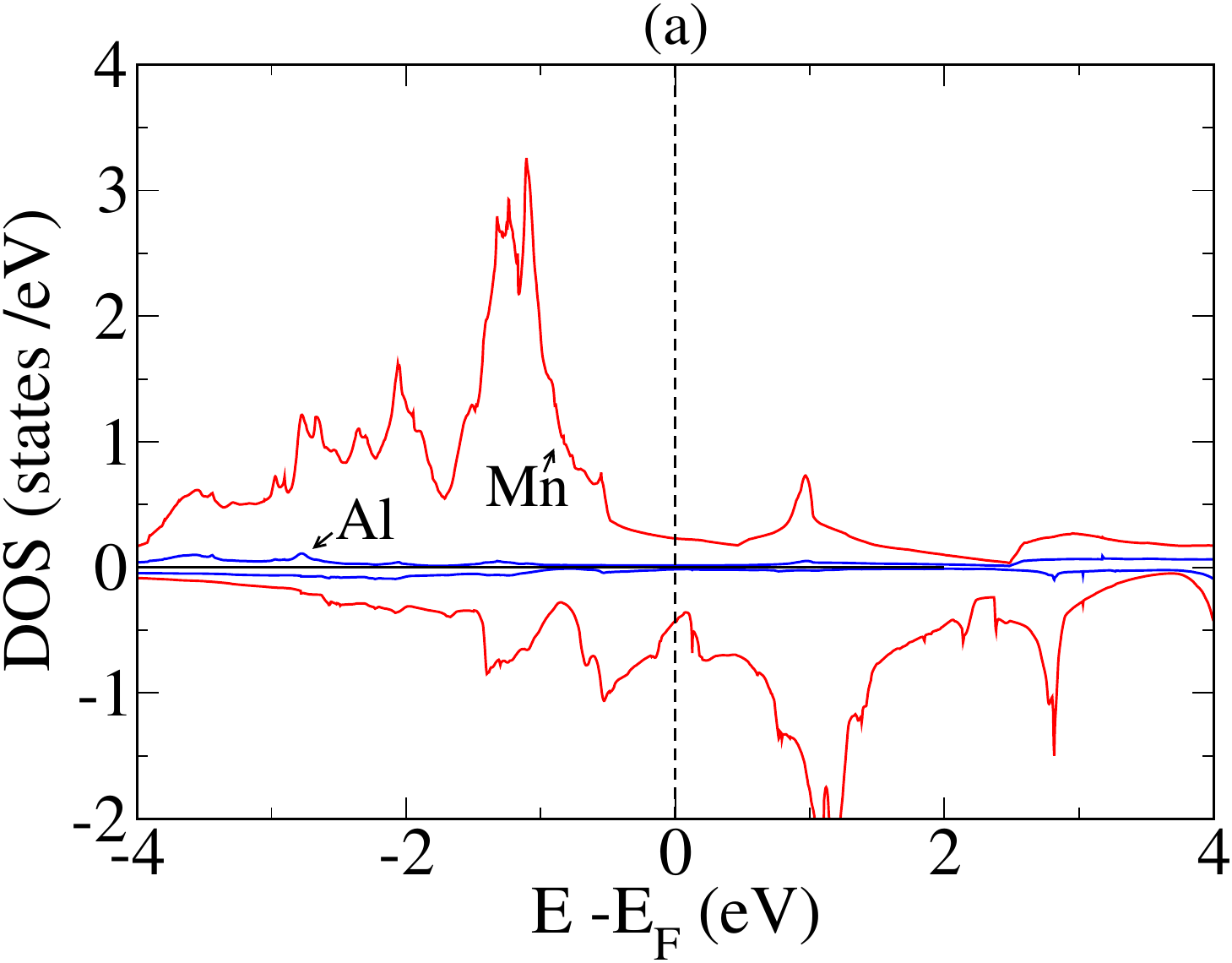}
\includegraphics[scale=0.275]{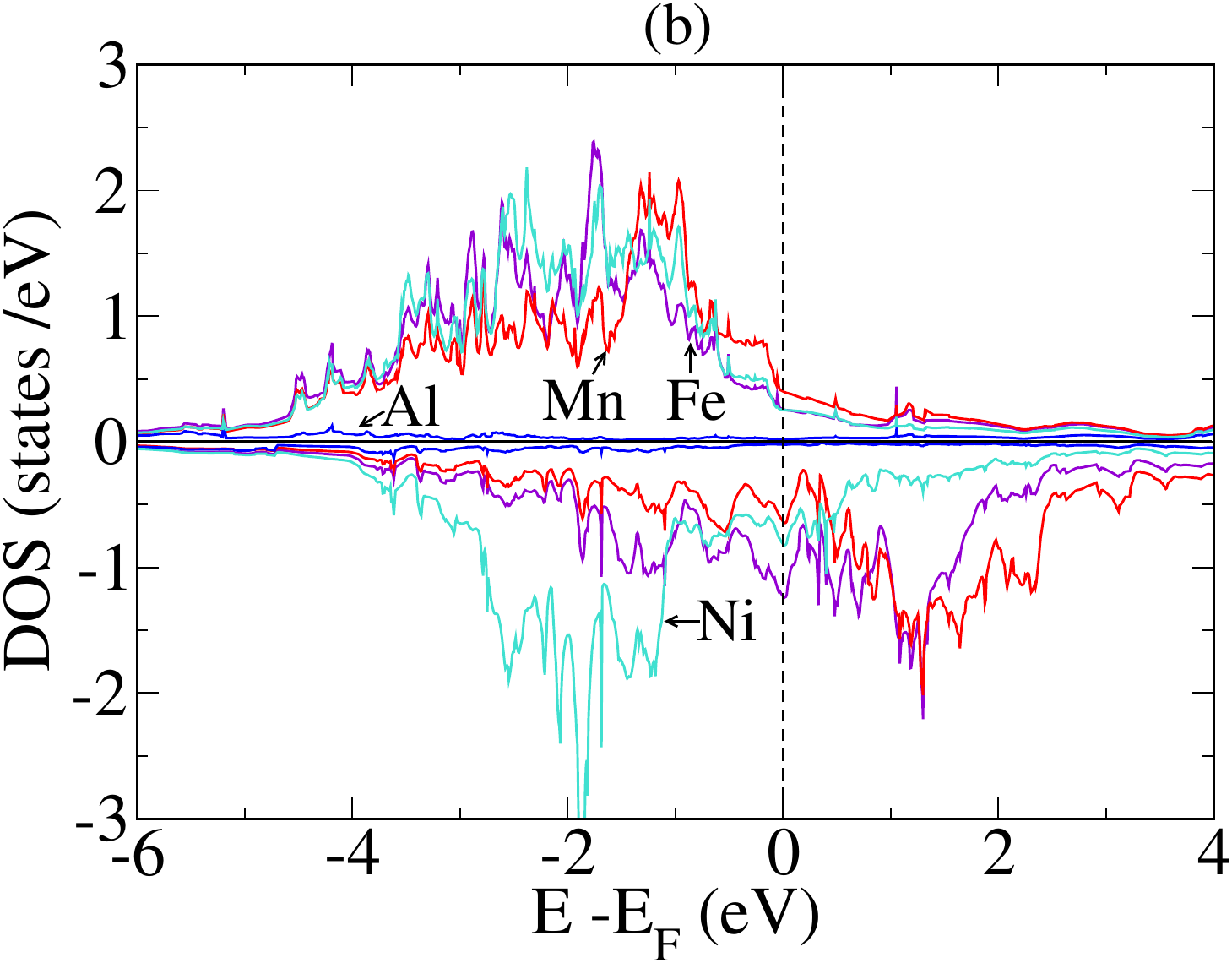}
\includegraphics[scale=0.275]{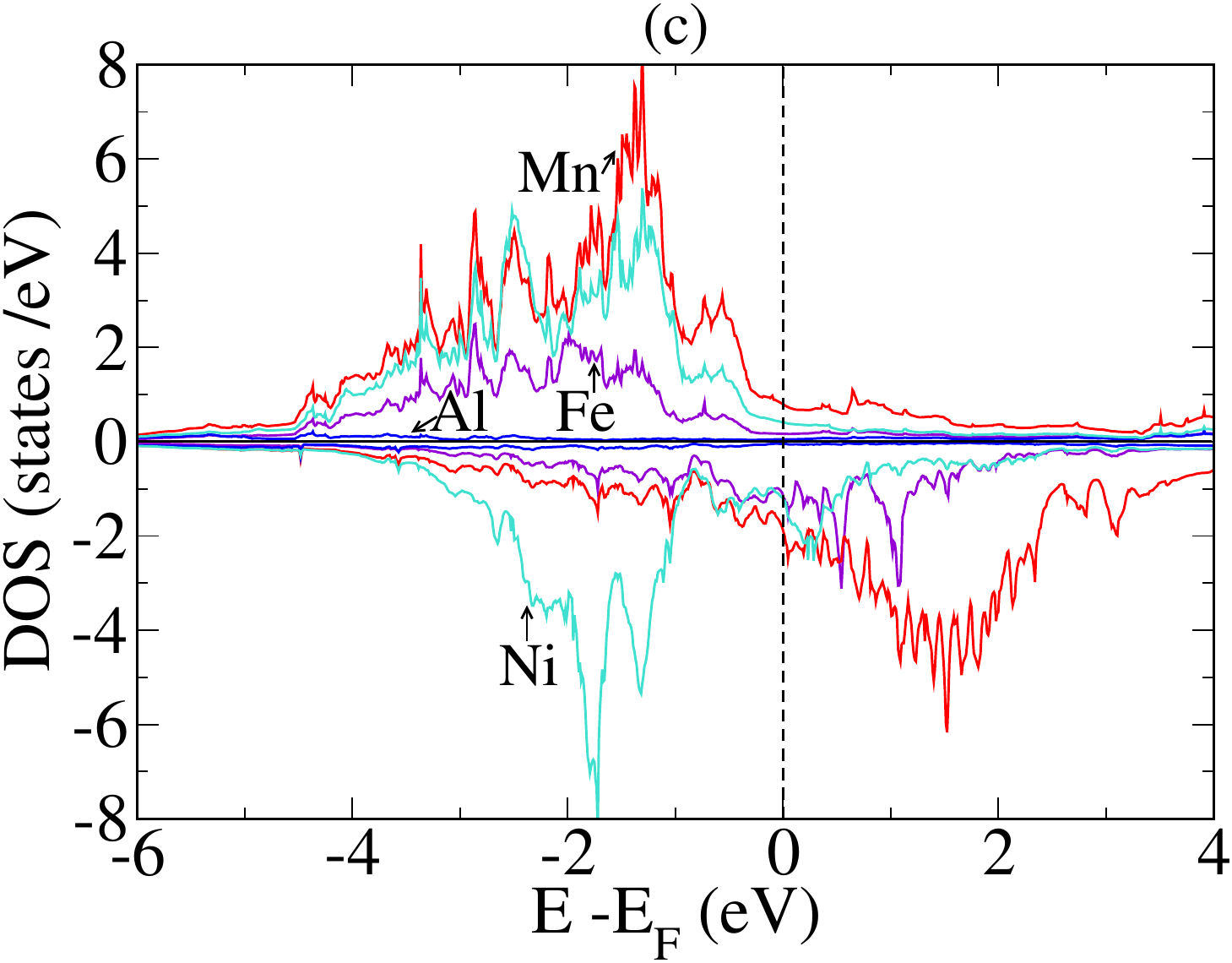}
\includegraphics[scale=0.275]{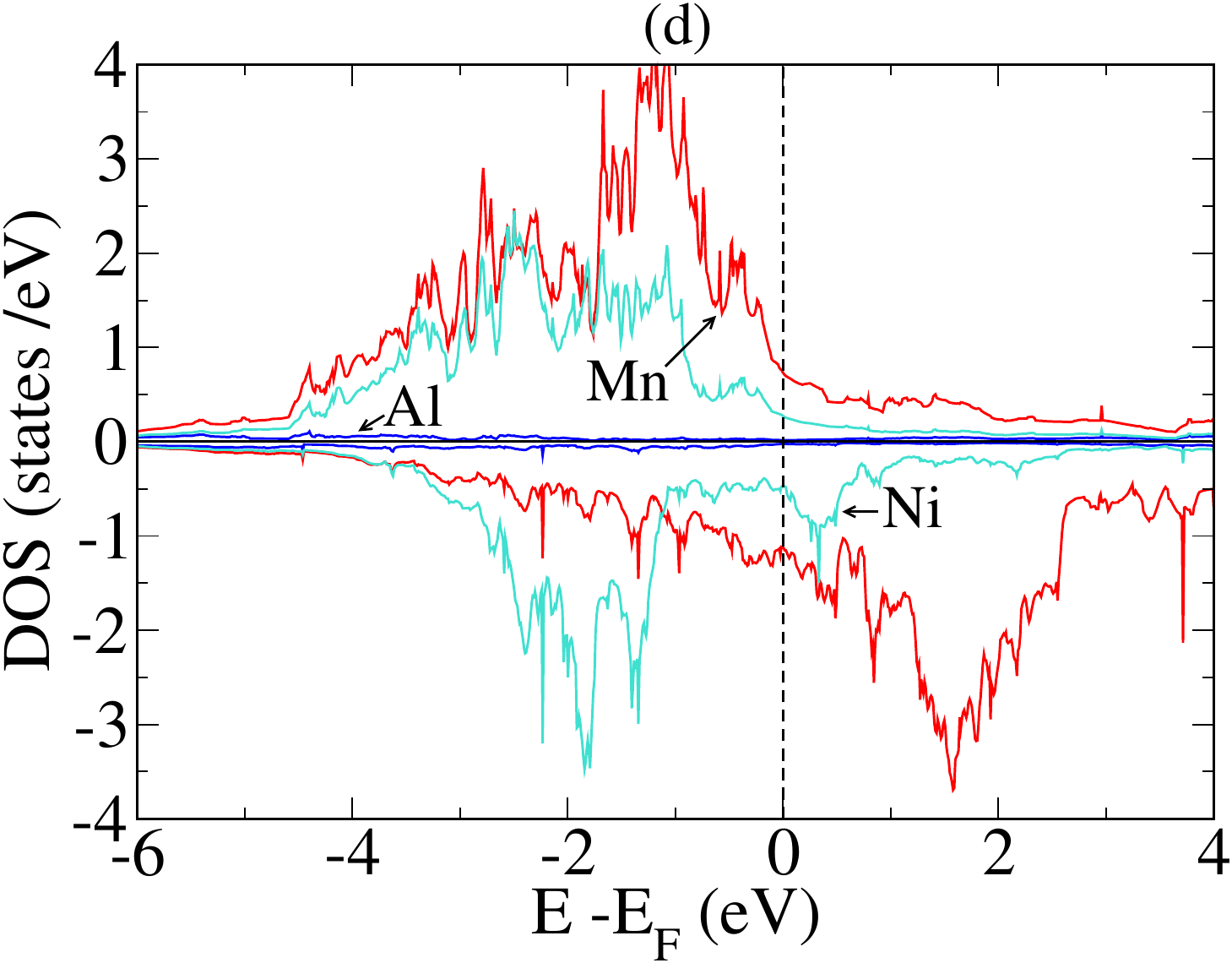}
\caption{Atom resolved electronic density of states (DOS) in (a) MnAl, (b) FeMnAlNi, (c) Fe$_{0.5}$Mn$_{1.5}$AlNi, and (d) Mn$_2$AlNi.}
\label{fig:dos}
\end{figure}

Figure \ref{fig:dos} shows the electronic density of states for (a) MnAl and (b) - (d) Fe/Ni substituted-MnAl obtained with spin-polarized GGA calculations. The DOS shows primarily Mn($3d$) states around the Fermi level with a minimal admixture of Al($sp$) states.
As expected for the $3d$-states, the spin-polarized calculations produce a magnetic state. The characteristics of the $3d$-states around the Fermi level change with Fe/Ni substitutions. Notably, varying the Fe content brings about intriguing modifications in the electronic structure. Initially, in the absence of Fe in MnAl (Fig. \ref{fig:dos} (a)) and Mn$_2$AlNi (Fig. \ref{fig:dos} (d)), the Fermi level lies away from the peak in the DOS. This condition is closely related to a lower formation energy, as demonstrated in Fig. \ref{fig:formation}, indicating enhanced chemical stability. When the Fe content reaches $25\%$, a minor peak emerges in the Mn-dos (Fig. \ref{fig:dos} (c)), which, however, remains negligible compared to the Fe and Mn-DOS peaks in Fig. \ref{fig:dos} (b). This behavior is consistent with the observed increase in formation energy with Fe content.  

The magnetic moment is susceptible to the Fe and Ni content. Substitutions affect the magnetic moment of individual magnetic atoms, thereby altering the net magnetic moment. All atoms, except Al, contribute to the magnetic moment. In the case of MnAl, each Mn atom exhibits a magnetic moment of 2.31 $\mu_B$, which is in excellent agreement with experimental findings \cite{KonoJPS58}. Effectively, the Mn spin magnetic moment increases with substitutions. For a Fe content of 25\% and a Ni content of 50\%, the net magnetic moment amounts to 12.53 $\mu_B$, albeit slightly decreasing with a further increase in Fe content. It is noted that the fractional amount of doping leads to a larger supercell. For 24\% Fe and 50\% doping, the supercell is four times larger than the pristine MnAl unit cell, as discussed in Section II. Therefore, the magnetic moment per unit cell has to be compared accordingly. Ni is crucial in enhancing the net moment by increasing the Mn moment and contributing to the overall magnetic moment. Fe also influences the net magnetic moment similarly, although its effect is not directly proportional to the Fe content. This discrepancy may arise from differences in the crystalline electric field. Despite the presence of orbital moments for all atoms, the disparity between the hard and easy axes is significant enough to induce desirable magnetic anisotropy through spin-orbit coupling.

\section{Magnetic anisotropy}
\begin{table}
\begin{ruledtabular}
\caption{The calculated spin moment ($\mu_s$) and orbital magnetic moment ($\mu_l$) (obtained with dense {\bf k}-mesh given in Section II) and comparisons with previous theory and experimental values. The total magnetic moments are given in per unit cell. Note that the unit cell of Fe$_{0.5}$Mn$_{1.5}$AlNi and Mn$_2$AlNi are four times, and FeMnAlNi is two times larger than MnAl. We present magnetization saturation density $J_s$ (in T) for better comparison. The partially Fe-substituted compositions have the same crystalline symmetry, while Mn$_2$AlNi has a higher symmetry space group orthorhombic no. 65. For FeMnAlNi, a distorted tetragonal structure was used.}
    \label{tab:moment}      
    \begin{tabular}{lllllll}
    Material/moments & Fe & Mn & Al & Ni & total & $J_s$\\
    \hline
    MnAl&&\\
      $\mu_s$   & - & 2.31 &-0.04 &- &2.32 & 1.04\\
      $\mu_l$ & - & ~0.03 & 0.00 &- &0.03\\
      Expt. & && & &2.31\footnote{Kono \cite{KonoJPS58}} & \\
      Theory & & & & &2.37\footnote{LAPW Ref.~[\cite{ParkJAP010}]} & \\
      \hline
      Fe$_{0.5}$Mn$_{1.5}$AlNi & &\\
       $\mu_s$   &2.38  & 3.00 & -0.07 & 0.65 & 12.53 & 1.48\\ 
       $\mu_l$ & 0.08 & 0.02 & 0.00 & 0.06 & 0.25 & \\
      
      \hline
      FeMnAlNi &\\
       $\mu_s$   &2.04  & 2.96 & -0.06 & 0.34 & 5.29 & 1.28\\ 
       $\mu_l$ & 0.05 & 0.02 & 0.00 & 0.02 & 0.08 & \\
       \hline
       Mn$_2$AlNi &&\\
       $\mu_s$ &  & 2.88 &-0.07&0.6 & 12.65 & 1.46\\
       $\mu_l$ & & 0.02& 0.00& 0.05 & 0.17 &\\
    \end{tabular}
    \end{ruledtabular}
\end{table}


\begin{table}
\begin{ruledtabular}
\caption{Comparison of calculated magnetic anisotropy constant ($K_u$ in MJ/m$^3$) with experiment. For Fe$_{0.5}$Mn$_{1.5}$AlNi and FeMnAlNi $[1,0,0]$ is easy and $[0,1,0]/[0,0,1]$ are hard axes for Fe$_{0.5}$Mn$_{1.5}$AlNi, while $[1,0,0]$ is hard and $[0,0,1]$ is easy axis for pristine MnAl. The values of $K_u$ relative to the [0,1,0] axis differ slightly depending on the choice of {\bf k}-points for FeMnAlNi. Mn$_2$AlNi is planar along $[1,0,0]$ and $[0,0,1]$-directions.}\label{aniso}  
\begin{tabular}{l l l l l}
$K_u$ &MnAl& FeMnAlNi &Fe$_{0.5}$Mn$_{1.5}$AlNi & Mn$_2$AlNi\\
\hline
This work  & 1.53\footnote{With {\bf k}-mesh $9\times9\times7$} &2.34\footnote{Distorted tetragonal structure with {\bf k}-mesh $10\times10\times10$, $K=E_{[100]}-E_{[001]}/V$, while $K=E_{[010]}-E_{[001]}/V = 1.61$ Mj/m$^3$}, 1.76\footnote{With {\bf k}-mesh $7\times10\times19$, 1.93 with respect to [0,1,0]} &1.26\footnote{With {\bf k}-mesh $4\times5\times9$, 0.58 with respect to $[0,0,1]$} & -0.68\footnote{With {\bf k}-mesh $4\times5\times9$, planar along $[1,0,0]$ and $[0,0,1]$}\\ 
& 1.46\footnote{With {\bf k}-mesh $13\times13\times14$} &$-$ & 1.34(0.45$^f$) &-0.57\\
&$-$  & $-$ & $-$& $-$\\
Theory& 1.5\footnote{LMTO-ASA Ref.[\cite{SakumaJPS94}]}, 1.53\footnote{LAPW  Ref.[\cite{ParkJAP010}]} &$-$ & $-$&$-$\\
& 1.67\footnote{LAPW Ref.[\cite{AlexanderPRB014}]}&$-$ &$-$ &$-$\\
Expt. &1.2\footnote{Thin film \cite{OoganeJJAS017}}, 1.7\footnote{\cite{KlemmerScripta95}} &$-$ & $-$&$-$
\end{tabular}
\end{ruledtabular}
\end{table}
Multiple test calculations were conducted using different sets of {\bf k}-points. The resulting $K_u$ values for all compounds are presented in Table \ref{aniso} for two converged sets of {\bf k}-points. It was observed that all considered compositions exhibited uniaxial anisotropy, with an easy axis aligned with the longest crystallographic direction. The value of $K_u$ was computed as $K_u=\big(E_{easy}-E_{hard}\big)/V$, where $E_{easy}$ and $E_{hard}$ represent the energies calculated for local spin moments along the easy and hard axes, respectively.

In the case of MnAl, the anisotropy was found to be uniaxial along the $[0,0,1]$ direction, confirming that $[1,0,0]$ and $[0,1,0]$ serve as the hard axes, while $[0,0,1]$ acts as the easy axis consistent with previous theoretical and experimental studies. Utilizing a {\bf k}-mesh size of either $9\times9\times7$ or $13\times13\times13$ yielded similar values of $K_u$ (see Table \ref{aniso}), the easy axis was determined to be $[1,0,0]$, with $[0,1,0]$ and $[0,0,1]$ acting as the hard axes in Fe$_{0.5}$Mn$_{1.5}$AlNi, except for Mn$_2$AlNi, where the anisotropy is planar. We note that the easy axis directions are defined relative to the primitive unit cell of the alloy supercells. For instance, [0,0,1] easy axis for MnAl corresponds to the $c$-axis of its unit cell, which is not the same as the $c$-axis of the alloy unit cell.
The size of the {\bf k}-mesh was adjusted proportionally to the lattice constants relative to the MnAl crystal structure.

\begin{table}
\begin{ruledtabular}
\caption{Magnetic anisotropy energy contributions (in meV) from different atoms in each composition. The value outside the bracket denotes the anisotropy energy contribution of the element inside the bracket. In Fe$_{0.5}$Mn$_{1.5}$AlNi, one Mn has negative while other remaining two have positive moments. Ni exhibits a dual nature i.e., positive or negative moments depending on the Mn doping concentration. \label{onsite}}
\begin{tabular}{l| l| l}
 MnAl & FeMnAlNi & Fe$_{0.5}$Mn$_{1.5}$AlNi\footnote{$4\times 5 \times9$}\\ 
 \hline
 0.478[Mn]   & ~1.112[Fe]    &   ~0.446[Fe] \\
 0.019[Al]   & ~0.43[Mn]    &  -0.027,~0.140[Mn] \\
          & ~0.00[Al]    &  -0.002[Al] \\
            &  -0.2[Ni]   &   ~0.304[Ni] \\
\end{tabular}
\end{ruledtabular}
\end{table}    

In the case of FeMnAlNi, $K_u$ increases by $\approx 56\%$ compared to MnAl. 
In the Fe$_{0.5}$Mn$_{1.5}$AlNi alloy, which corresponds to a 25\% Fe-doped composition, the value of $K_u$ decreases by approximately 20\%. This suggests that reducing the Fe content enhances the chemical stability, but it diminishes the value of $K_u$ in the Ni-doped compositions. In fact, without Fe, the Mn$_2$AlNi compound becomes planar within the $ac$-plane. A more detailed understanding can be obtained through the site-resolved anisotropy, which is calculated by considering the onsite spin-orbit energy difference between the easy and hard axes for each atomic site.

According to Table \ref{onsite}, Mn exhibits uniaxial behavior in all cases except for the 25\% Fe-doped alloy, where one Mn atom becomes planar. Fe consistently exhibits uniaxial behavior and significantly contributes to $K_u$, consistently surpassing that of Mn. On the other hand, as expected for the $sp$ orbital system, Al contributes the least to the magnetocrystalline anisotropy energy.

{\section{Machine learning}
\subsection{Formation energy and saturation magnetization density}

\begin{figure}
    \centering
    \includegraphics[width=0.9\linewidth]{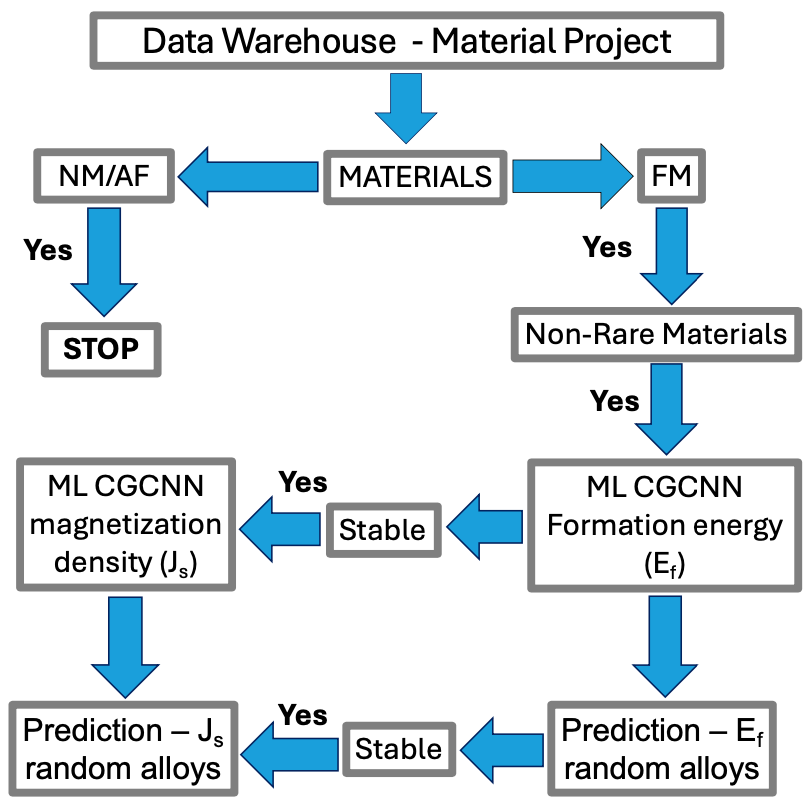}
    \caption{Flowchart showing the ML-CGCNN workflow: Database collection, ML training on DFT parameters, and predicting new magnetic alloys compositions.}
    \label{fig:flowchart}
\end{figure}
Figure 5 shows a pipeline for extracting and processing crystal structure properties from the Materials Project\cite{JainAPL013}. The scraped dataset consists of 153K crystal structures with electronic and magnetic properties of magnetic and non-magnetic materials. For our purpose, we want to train only the ferromagnetic materials properties appropriate for Mn and Al binary alloy predictions. We down-select the ferromagnetic 50K materials, consisting of rare-earth and without rare-earth elements. The DFT calculated saturation magnetic density ($J_s$) of rare-earth-based materials in the Material project excludes the $4f$-orbitals in the frozen-$4f$ orbital. To better predict $J_s$ for binary alloys, we filtered out the rare-earth-based magnetic databases for ML. To be consistent, we only utilized ferromagnetic materials without rare-earth elements for ML training. In our calculations, we have 47K non-rare-earth ferromagnetic materials, which we split into 6:2:2 as training, validation, and test datasets in the crystal graph convolution neural network (CGCNN) algorithm\cite{XiePRL018}. The model has a validation accuracy the MAE of 0.098 eV/atom which is similar to that found in \cite{XiePRL018} with CGCNN and \cite{Kirklin2015} obtained with the high-throughput calculations in Open Quantum
Materials Database. In Figure \ref{ML_test_performance}, we compare $E_f$ between the ML prediction and DFT results in the test dataset.

For $J_s$ ML training, we used 34K non-rare-earth element-based stable compositions, including the magnetic moment in the range 1.6 - 30 $\mu_B$ per formula unit for better ML training of realistic magnetic materials. The data is split into train, validation, and test sets as a 6:3:1 ratio, which produces validation accuracy the MAE of 0.061 T. Fig. \ref{ML_test_performance} shows the comparison between ML prediction and DFT results for $J_s$ in the bottom panel.

\begin{figure}   \includegraphics[width=0.925\linewidth]{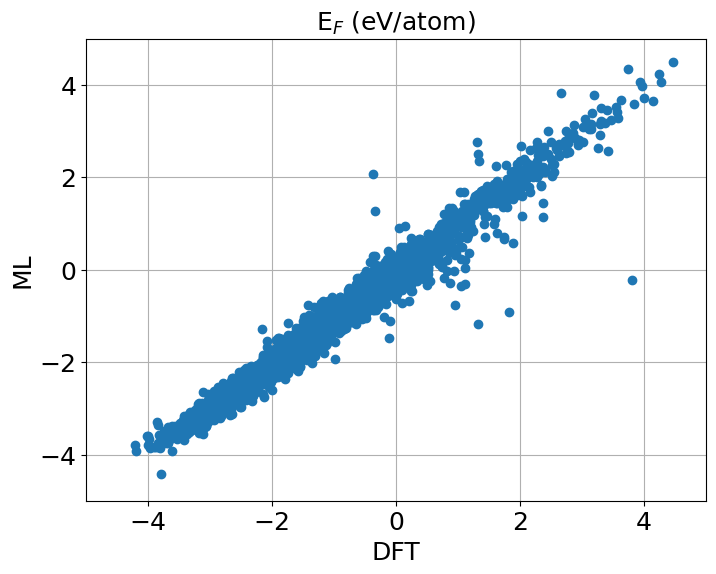}
\includegraphics[width=0.95\linewidth]{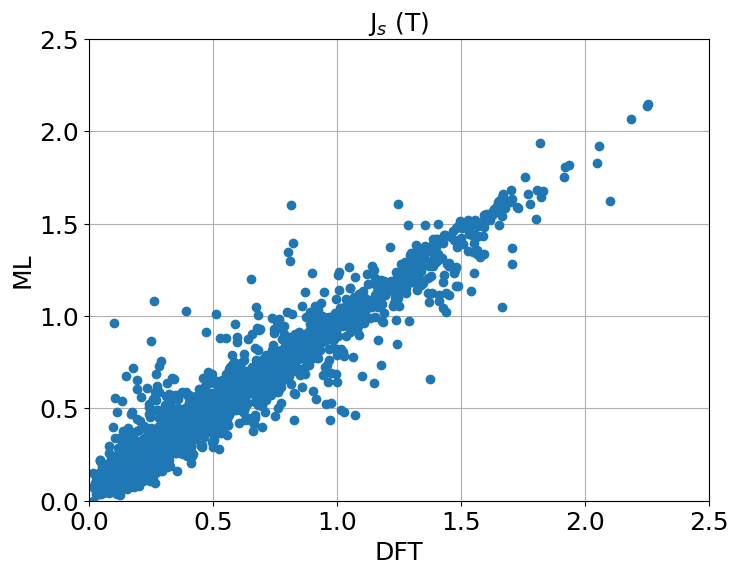}
\caption{Comparison of formation energies in {\it top} and saturation magnetization density {\it bottom} obtained with DFT and CGCNN-ML prediction in the test dataset.}
    \label{ML_test_performance}
\end{figure}

{\subsubsection{Machine learning predictions on alloys}
We generate random alloys by replacing Mn with Fe and Al with Ni, building a $4\times4\times4$ supercell containing 256 atoms. The total number of possible resulting compositions is 16384. The unit cell volume is scaled according to the covalent radii of the dopants to match two limits: undoped MnAl and fully-doped FeNi.
The ML-predicted formation energy is shown in Fig.~\ref{ML_formation} as a function of Fe and Ni concentrations. The heatmap indicates that increasing Fe increases the stability for small Ni doping concentrations. While the alloy compositions are stable, the stability decreases with higher Ni-doping concentrations. The predicted formation energy of MnAl, FeAl, MnNi, and FeNi are respectively -0.28, -0.48, -0.12, and -0.08 eV/atom which are consistent with DFT values -0.27, -0.33, -0.04, and -0.07  eV/atom in Material project\cite{JainAPL013}. The predicted formation energy for 50\% doped MnAl with Fe and Ni -0.28 eV/atom is slightly less than MnAl, which qualitatively agrees with our DFT calculations. The small numerical discrepancy is attributed to a mismatch in the lattice constants. The formation energy decreases with simultaneous increase in Fe and Ni concentrations.
\begin{figure}
    \centering
    \includegraphics[width=1.0\linewidth]{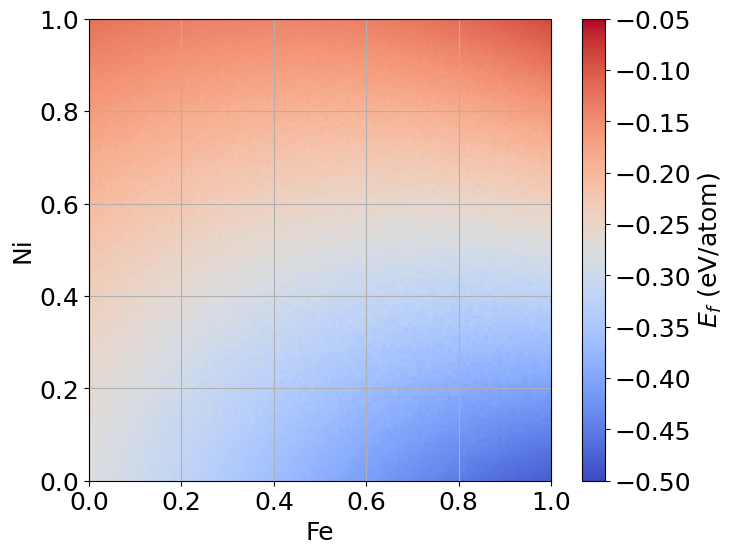}
    \includegraphics[width=1.0\linewidth]{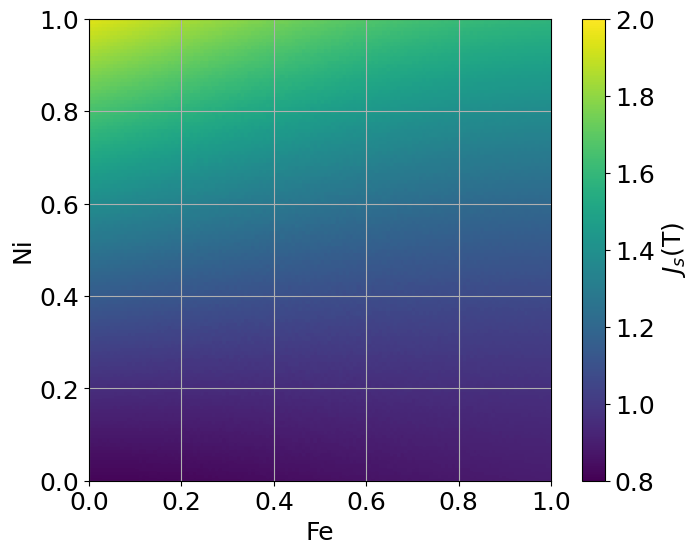}
    \caption{Heat-map showing the variation of formation energy encoded in color map as function of dopants Fe and Ni concentrations {\it top} and saturation magnetization density $J_s$ in {\it bottom}.}
    \label{ML_formation}
\end{figure}

The trend in saturation magnetization density is opposite comparing with the formation energy. Ni tends to increase it, while Fe has the opposite effect. The predicted $J_s$ are 0.81, 0.89, 1.96, and 1.59 T for MnAl, FeAl, MnNi, and FeNi respectively. For MnFeAlNi (which amounts 50\% doping) the predicted $J_s$ is 1.17 T. The ML predictions show that Fe reduces the $J_s$ while Ni does the opposite consistent with DFT calculations as given in Table \ref{tab:moment}. 

\section{Micromagnetic prediction}
The coercivity and hysteresis loop of the pristine and doped MnAl are investigated using micromagnetic simulations ($\mathrm{mumax}^3$)\cite{mumax1, mumax2, mumax3, mumax4}. While micromagnetic analysis typically overestimates the absolute values of coercivity, it provides valuable insights into the trends and relative changes in coercivity caused by elemental substitutions\cite{BrownparadoxRevModPhys45, AharoniRevModPhys62, HartmannPRB87}.

The magnetodynamics of magnetic material is governed by the nonlinear partial differential equation for the spatio-temporal magnetization vector ${\bf M}({\bf r},t)$, known as the Landau–Lifshitz–Gilbert (LLG) equation \cite{Brown1963,GilbertIEEE04}:

\begin{eqnarray}
\dfrac{\partial {\bf M}}{\partial t} = \dfrac{\gamma}{1+\alpha^2}{\bf M}\times {\bf H_{\mathrm{eff}}} - \dfrac{\alpha\gamma}{1+\alpha^2} {\bf M} \times\bigg[{\bf M}\times {\bf H_{\mathrm{eff}}}\bigg],
\label{GL}
\end{eqnarray}

where $\gamma$ is gyromagnetic ratio, $\alpha$ damping constant, and ${\bf H_{\mathrm{eff}}}$ is effective magnetic field.
The first term in Eq.~\eqref{GL} describes precession, the gyroscopic motion of the magnetic moments around the effective magnetic field, while the second damping term captures the relaxation of the moments towards equilibrium. The damping parameter $\alpha$ is set to $0$ for time-independent scenarios, such as hysteresis loop calculations.

The effective magnetic field ${\bf H_{\mathrm{eff}}}$ consists of several contributions: the externally applied field ${\bf H}$, the demagnetizing field ${\bf H_{\mathrm{dm}}}$ arising from the long-range dipolar interactions, the exchange field ${\bf H_{\mathrm{ex}}}$ derived from the Heisenberg model, and the uniaxial anisotropy field ${\bf H_{a}}$.

The demagnetizing field ${\bf H_{\mathrm{dm}}}$ plays a crucial role in determining the shape anisotropy of the MnAl nanostructures, which can significantly influence the coercivity and the shape of the hysteresis loop. The exchange field ${\bf H_{\mathrm{ex}}}$, characterized by the exchange stiffness constant $A_{\mathrm{ex}}$, promotes the alignment of neighboring magnetic moments and affects the domain wall width and the smoothness of the magnetization transitions.

The uniaxial anisotropy field ${\bf H_{a}}$ is determined by the $K_u$ and the anisotropy direction ${\bf u}$. In MnAl, the magnetocrystalline anisotropy arises from the combined effect of spin-orbit coupling of Mn $3d$-states and crystalline electric field in the ordered tetragonal L1$_0$ structure, leading to a strong easy-axis anisotropy along the c-axis. 

The key material parameters $K_u$, $M_s$ ($J_s$), and $A_{\mathrm{ex}}$ capture the essential physics of MnAl and enable the micromagnetic simulations of intricate magnetic structures. By systematically varying these parameters and studying their effect on the coercivity and hysteresis loop, a deeper understanding of the magnetization reversal processes and the role of material modifications in tailoring the magnetic properties of MnAl for specific applications, such as high-performance permanent magnets and magnetic recording media is gained.

\begin{table}
\begin{ruledtabular}
\caption{Magnetic parameters used in the micromagnetic simulations and predicted coercivity ($\mathrm{H}_c$). $A_{ex}$ was taken from the experiment and assumed constant. $J_s$ and $K_u$ were obtained from DFT calculations for each material, using the largest {\bf k}- mesh size.\label{micromag}}
\begin{tabular}{l|lll}
 parameters & MnAl & FeMnAlNi & Fe$_{0.5}$Mn$_{1.5}$AlNi\\ \hline
 $A_{ex}$ (pm) & 7.6 & 7.6 & 7.6 \\
 $K_u$ (MJ/m$^3$) & 1.53 & 2.34 & 1.26 \\
 $J_s$ (T) & 1.04 & 1.28 & 1.48 \\
 $\mathrm{H}_c$ (T) & 0.47 & 0.55 & 0.23 \\
\end{tabular}
\end{ruledtabular}
\end{table}    

As reported earlier by us in Ref.~[\onlinecite{bhandariML023}], micromagnetic simulation overestimates the $H_c$ of permanent magnets by about a factor of 5 without grain boundary engineering due to Brown paradox\cite{BrownparadoxRevModPhys45, Brown1963}.
To compute a reasonable estimate of coercivity, a mesh size of $128^3$ (a magnetic cuboid) was chosen with each microcell size of $1\,$nm$^3$. The direction of the uniaxial magnetic anisotropy was taken to be uniform across grains of width $20\,$nm, and each grain was a random region chosen by Voronoi tesselation with the direction of the anisotropy also in a uniformly random direction. Inter-grain coupling was taken to be $10\%$ lower across the grains than the intra-grain exchange stiffness constant, $A_{ex}$, value of $7.6\,$pm within the grains. All magnetic parameters used in the micromagnetic simulations are given in Table~\ref{micromag}, except the inter-grain coupling and anisotropy constant, which were varied.

For the computation of the hysteresis loop, a uniform magnetic field was applied at $5\,$T, which was then reduced in steps of $0.01\,$T until $-5\,$T was reached and the average reduced magnetization was measured (the large variation in $M_s$ makes the reduced magnetization where the material's magnetization is put as a ratio against the total possible magnetization easier for visualization). The resulting path was smoothed with a weighted average to adjust for the relative discreteness of the mesh compared to real materials. Moreover, due to the symmetric properties of this magnetic model, the other branch of the hysteresis loop was then computed by reflection across the origin. The hysteresis loop is visualized in Fig.~\ref{fig:coercivity}, and the coercivities of corresponding compositions are listed in Table~\ref{micromag}.

The na\"ive upper limit of $2K_u / M_s = H_c$ for coercivity closely tracks the material limitations evident in the detailed micromagnetic computations despite overestimating it due to the Brown paradox \cite{BrownparadoxRevModPhys45}. However, the Brown paradox also causes the micromagnetic simulations to overestimate the actual experimental coercivity of MnAl $0.27\,$T, indicating that the ratios in Table~\ref{micromag} are more important than the exact values obtained \cite{hc1}. The introduction of Fe increases $K_u$, decreasing it slightly in Fe$_{0.5}$Mn$_{1.5}$AlNi. However, the additional increase in $M_s$ due to the increase in $\mu_s$ reduces the coercivity to the lowest of the materials considered. The simple trend apparent in the table suggests increased iron content in the unit cell may further increase the $K_u$, leaving the $M_s$ approximately constant, suggesting that Fe$_{1.5}$Mn$_{0.5}$AlNi would have elevated $H_c$, requiring further material investigation.

\begin{figure}
\hspace{-0.9cm}\includegraphics[scale=0.35]{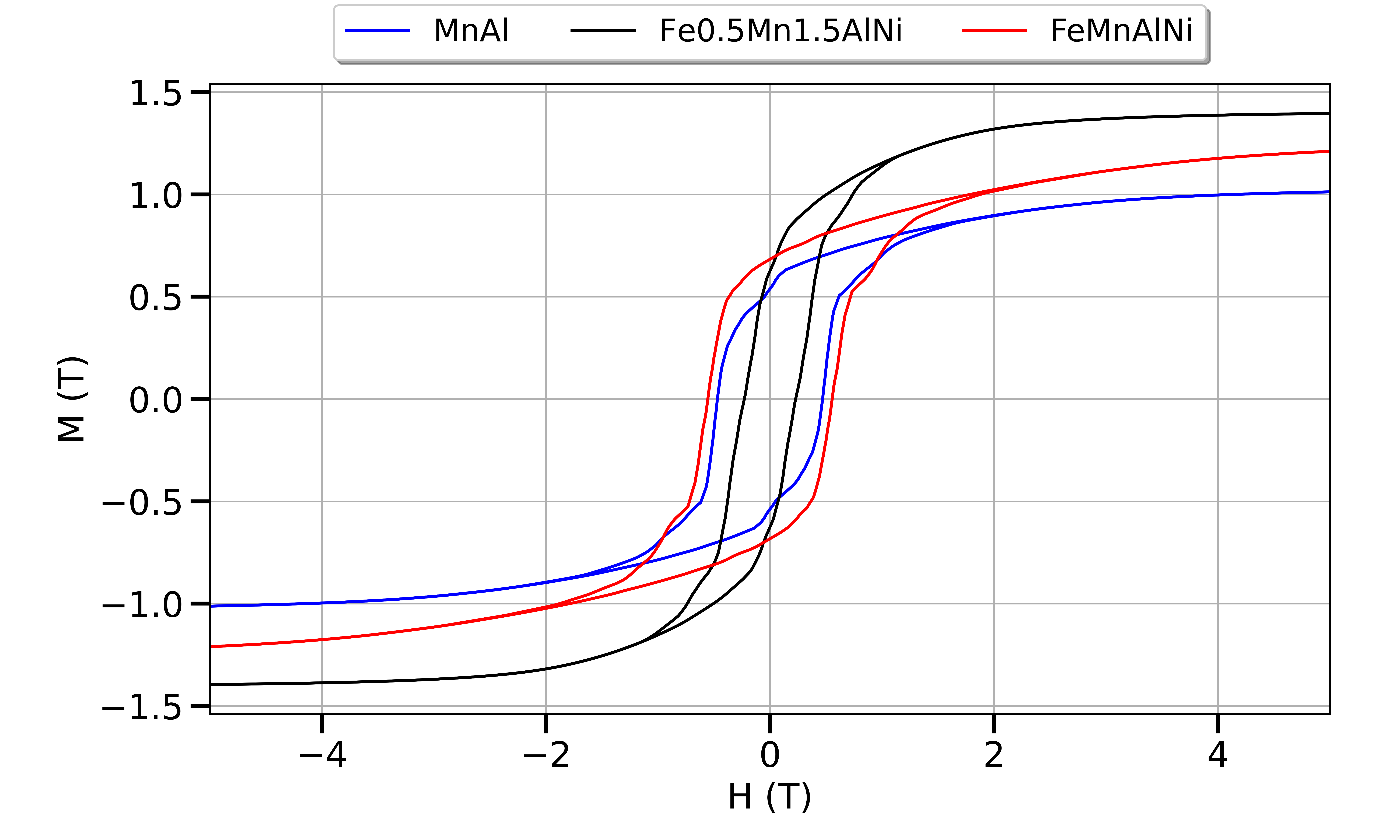}
\caption{The hysteresis loops of pristine and doped MnAl displaying the micromagnetically simulated normalized magnetization along the vertical axis and applied magnetic field $H$ along the horizontal axis.}
\label{fig:coercivity}
\end{figure}

\section{Conclusion}
Employing first-principles calculations and ML-CGCNN predictions, this work investigates how the substitution of Fe and Ni in the L$1_0$ type $\tau$-phase of MnAl affects its magnetic properties. The presence of these dopants has an impact on the stability and electronic characteristics. Adding Fe yields intriguing properties dependent on the crystal structure when the Ni-content is held constant. Specifically, the formation energy decreases in distorted tetragonal and monoclinic structures, reducing Fe content. ML predictions suggest that $E_f$ decreases with Fe for small Ni-dopings, while it increases with increasing Ni. All proposed compositions of the L$1_0$ type are mechanically and dynamically stable. Notably, a higher percentage of Fe enhances the magnetic anisotropy and magnetic moments. Surprisingly, the $E_f$ increases reducing the stability with simultaneously Fe and NI dopings, while the $J_s$ increases. Remarkably, the predicted $H_c$ is improved for the equiatomic phase obtained by simultaneous doping of Fe and Ni. Our study provides insight for stabilizing the MnAl $\tau$ L$1_0$ phase, developing $H_c$, and enhancing its magnetic properties by substituting Fe and Ni.

\acknowledgements{This work is supported by the Critical Materials Institute, an Energy Innovation Hub funded by the U.S. Department of Energy, Office of Energy Efficiency and Renewable Energy, Advanced Manufacturing Office. The Ames National Laboratory is operated for the U.S. Department of Energy by Iowa State University of Science and Technology under Contract No. DE-AC02-07CH11358.}

\appendix
\section{Elastic stiffness tensor}\label{appendix}
In this appendix, the calculated elastic constant tensor for all the considered MnAl alloys is presented. For pristine MnAl it is given in (\ref{stiffMnAl})
\begin{equation}
C_{ij}=\begin{bmatrix}
292.795 &  43.09    & 89.18   &  0.000     & 0.00     &0.00\\
43.09   &  292.79   & 89.18   &  0.00      &0.00      &0.00\\
89.18   &  89.18    & 165.06  &  0.00      &0.00      &0.00\\
0.00    &   0.00    & 0.00    & 119.78     &0.00      &0.00\\
0.00    &   0.00    & 0.00    &  0.00      &119.780   &0.00\\
0.00    &   0.00    & 0.00    &  0.00      &0.00      &75.75\\
\end{bmatrix}.\label{stiffMnAl}
\end{equation}
It definitive positive and satisfies the criteria for mechanical stability of MnAl.
The calculated elastic constant tensor for FeMnAlNi is given below \ref{stiffFeMnAlNi}.
\begin{equation}
C_{ij}=\begin{bmatrix}
207.76   & 123.64     &105.41   & 0.00    &  0.00    & -0.39\\
123.64   &  290.69    & 17.15   & 0.00    &  0.00    & -5.81\\
105.41   &   17.15    & 236.80  &  0.00   &  0.00    &  7.61\\
0.00     &   0.00     & 0.00    &  28.40  &  2.33    &  0.00\\
0.00     &   0.00     & 0.00    & 2.33    &  115.13  &  0.00\\
-0.39    &   -5.81    & 7.61    &   0.00  &    0.00  &   92.22\\
\end{bmatrix}\label{stiffFeMnAlNi}
\end{equation}

Similarly, for Fe$_{0.5}$Mn$_{1.5}$AlNi (\ref{stiffFexMn3x}), the elastic constant tensors are

\begin{equation}
C_{ij}=\begin{bmatrix}
193.55   & 114.87     &100.31      &0.00      &0.00      & -2.32\\
114.87   & 272.15     &24.67       &0.00      & 0.00     & -1.66\\
100.31   &  24.67     &213.36      &0.00      &0.00      &  1.17\\
0.00     &    0.00    &  0.00      &41.58     & 0.08     &  0.00\\
0.00     &    0.00    &  0.00      & 0.08     &116.72    &  0.00\\
-2.324   &   -1.66    &  1.17      & 0.00     & 0.00     &  98.71\\
\end{bmatrix}. \label{stiffFexMn3x}
\end{equation}
For Mn$_{2}$AlNi (\ref{stiffMn2}), the diagonal elements are similar
\begin{equation}
C_{ij}=\begin{bmatrix}
193.15   & 111.18    &  107.63     & 0.00     & 0.00      &  0.00\\
111.18   & 221.44    &   15.33     &  0.00    &  0.00     &   0.00\\
107.63   &  15.33    & 226.37      &  0.00    &  0.00     &   0.00\\
0.00     &  0.00     &  0.00       & 54.44    & 0.00      &  0.00\\
0.00     &  0.00     & 0.00        & 0.00     & 115.64    &  0.00\\
0.00     &   0.00    &  0.00       & 0.00     &  0.00     & 106.71\\
\end{bmatrix}.\label{stiffMn2}
\end{equation}

\bibliography{MnAl, micromagnetic}
\end{document}